\documentclass[aps,twocolumn,amsmath,amssymb,floatfix]{revtex4}
\usepackage{epsfig,psfrag,subfigure}
\usepackage{graphicx,psfrag,subfigure}
\usepackage{color}
\begin{document}
\title{Cooper pair tunneling in junctions of singlet
  quantum Hall states and superconductors}
\author{Eun-Ah Kim}
\author{Smitha Vishveshwara}
\author{Eduardo Fradkin}
\affiliation{Department of Physics, University of Illinois at
Urbana-Champaign, 1110
West Green Street , Urbana, IL  61801-3080, USA}
\date{\today}

\newcommand{\D}{\displaystyle}
\newcommand{\noi}{\noindent}
\newcommand{\F}{{\textrm F}}
\definecolor{purple}{rgb}{0.59,0.04,0.57}
\definecolor{green}{rgb}{0.31,0.46,0.29}
\definecolor{red}{rgb}{0.62,0,0.15}

\begin{abstract}
We propose tunnel junctions of a Hall bar and a superconducting
lead, for observing Cooper-pair tunneling into singlet fractional
quantum Hall edge states.
These tunnel junctions
provide a natural means of extracting precise information of the
spin polarization and the filling factor of the state. The low energy regime of one of the set-ups
is governed by a novel quantum entangled fixed point.
\end{abstract}
\maketitle



Junctions of fractional quantum Hall (FQH) states are unique tools to
investigate fascinating and unexpected aspects of Quantum Mechanics. In this
paper we discuss the physics of {\em singlet} FQH/superconductor (SC)
junctions. Tunneling in these junctions involve Cooper pairs which either
tunnel as a whole, or split into their constituent electrons which, if
quantum coherence is not spoiled, must retain the spin-singlet nature of the
Cooper pair and thus show quantum correlations in apparently separate
physical systems. 

 The set-up we propose offers a direct means of probing the transition from 
spin-polarized to singlet QH (SQH) states observed for certain
filling factors, such as $\nu=2/5, \nu=2/3$, driven either by
application of hydrostatic pressure or by tilted magnetic
fields~\cite{cho98,leadley97,kronmuller98,eisenstein90}. Furthermore,
via enhanced tunneling conductance,
it can provide precise information of effective filling
factors by filtering out SQH states as a function of magnetic field.
The beauty of this setup lies in the fact that
the tuning of appropriate energy scales can
turn the tunnel junction into an entanglement device wherein two disconnected
QH 
droplets extract entangled electron pairs from the SC lead.
Cooper pair tunneling and splitting are a focus of interest in the context of quantum information 
in solid state systems such as quantum dots and
nanotubes~\cite{entanglement}.
In principle, the SC/SQH junctions discussed here have the advantage of bypassing
fabrication issues
present in these other set-ups, particularly with regards to the control of the contact
at the junction.  As we shall see, the low energy physics of the SQH/SC junctions
is governed by a quantum entangled fixed point (QEFP) which links together
the SC lead and two SQH edge states.

\begin{figure}[ht!]
\centering
\psfrag{Normal}{\scriptsize N}\psfrag{SC}{\scriptsize SC}\psfrag{QH}
{\scriptsize SQH}
\subfigure[]
{\includegraphics[width=.3\textwidth]{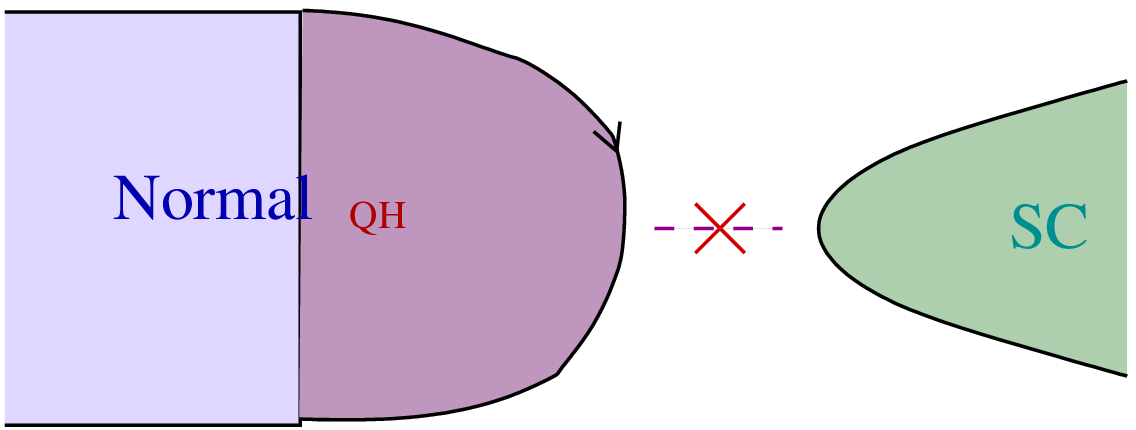}}\\
\psfrag{Normal}{\scriptsize N}\psfrag{SC}{\scriptsize SC}
\psfrag{QH}{\scriptsize SQH}
\subfigure[]
{\includegraphics[width=.3\textwidth]{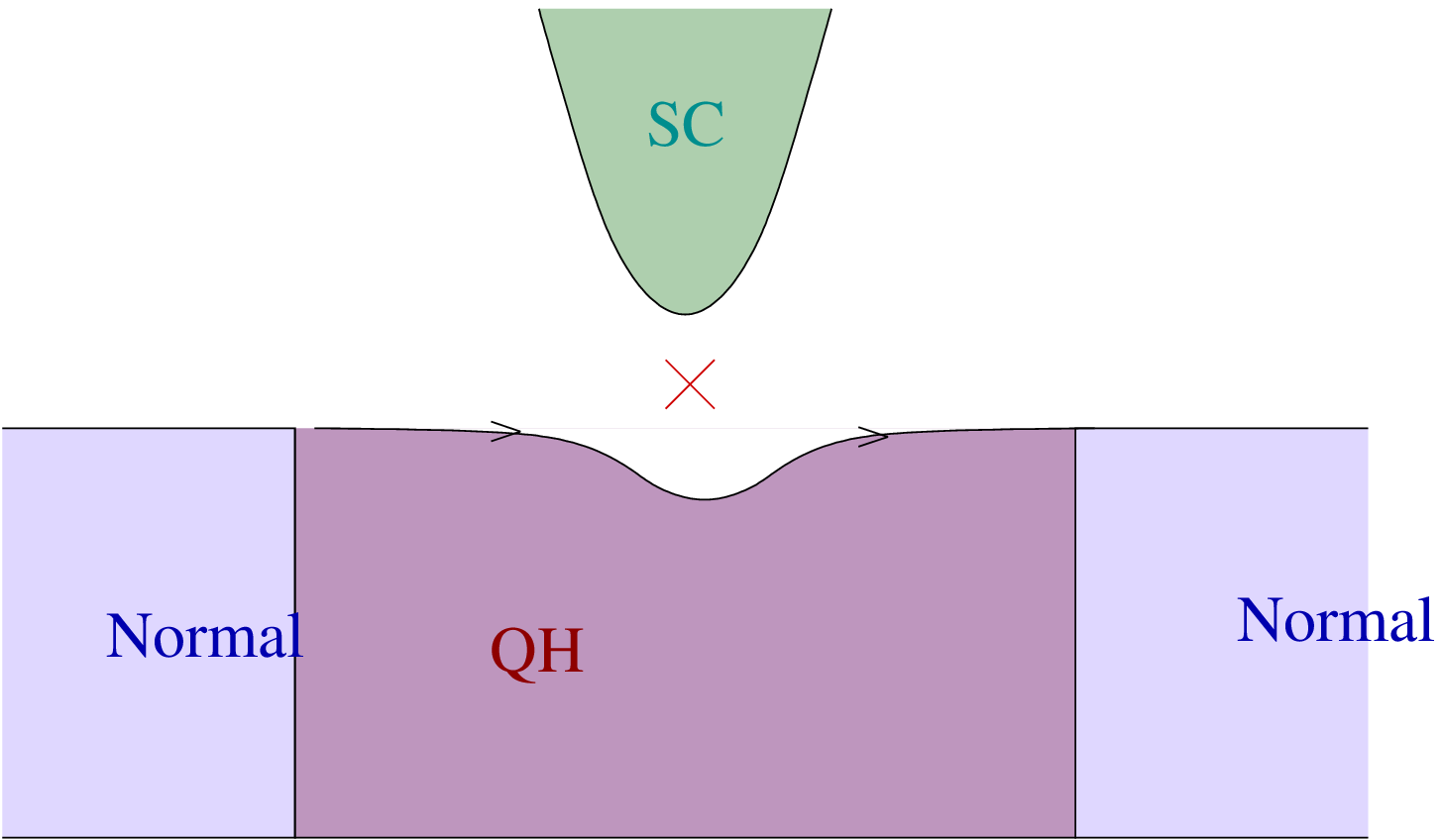}}\\
\psfrag{Normal}{\scriptsize N}\psfrag{SC}{\scriptsize SC}
\psfrag{QH}{\scriptsize SQH}
\subfigure[]
{\includegraphics[width =.3\textwidth]{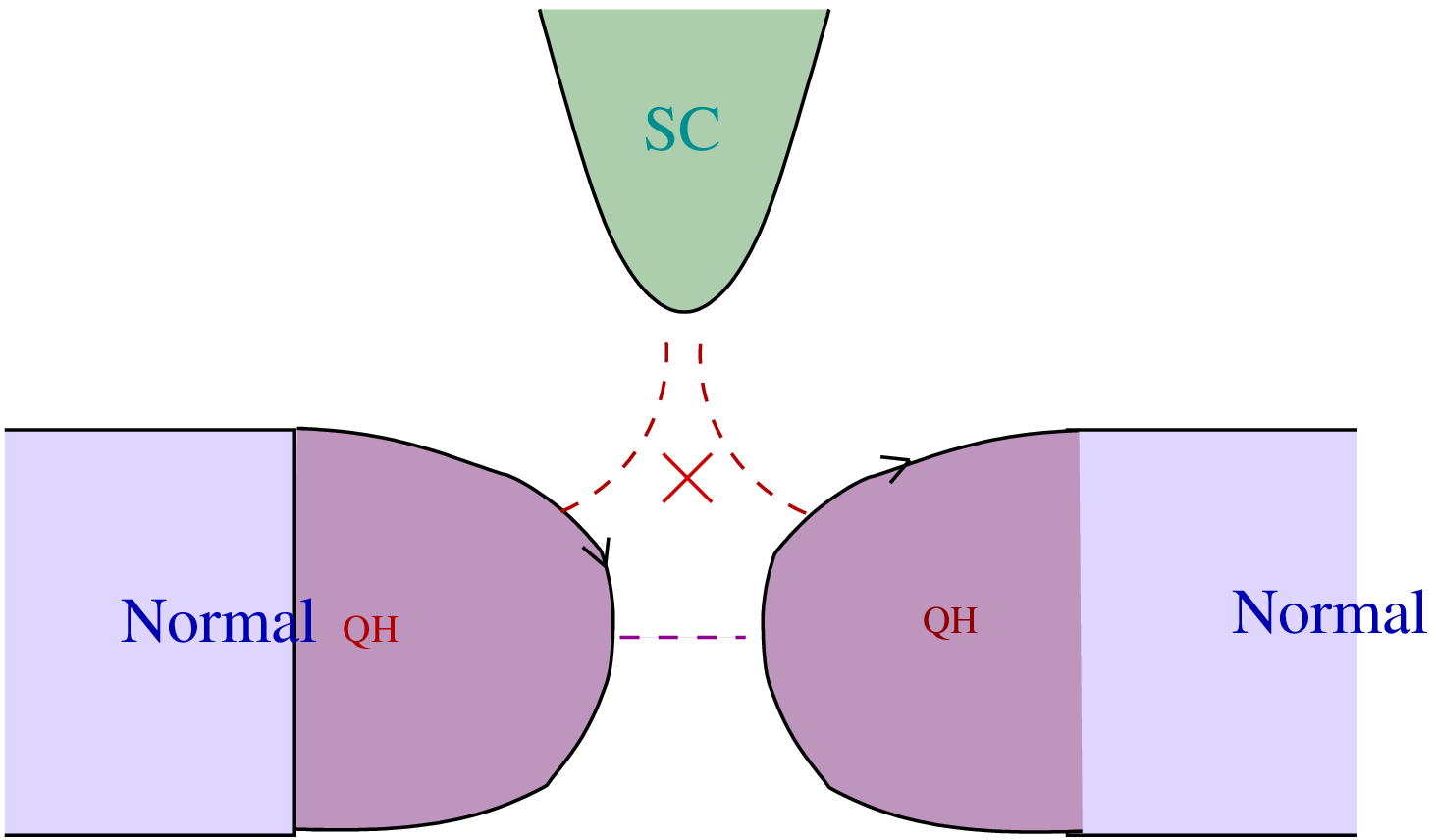}}
\caption{Cooper pair tunneling from a SC to SQH state. 
(a) Set-up I, with one SQH and one SC lead;
Set-up II, 
a SQH bar coupled to a SC lead in the (b) UV and (c) IR regimes. Notice the splitting of the SQH bar in (c).}
\label{fig:setup}
\end{figure}

We are interested in tunneling of Cooper pairs from singlet-paired SC 
electrodes  into  SQH
states sketched in Fig.\ref{fig:setup}.
So as to observe purely Cooper pair
tunneling, and no Bogoliubov quasiparticle excitations, we assume
all the important energy scales are smaller than the superconducting
gap energy. 
Notice  that the physics here is markedly different from that of conventional superconductor-insualtor-normal
(S-I-N) junctions, since although all of the QH bulk excitations are gapped, the QH droplet supports
gapless chiral edge excitations that provide a non-dissipative channel for conduction. 
Thus far, theoretical treatments of such a set-up have been relatively 
few \cite{fisher94,moore02} as it requires a superconductor
in high magnetic 
fields. However, experiments using NbN films
as SC leads in contact with QH systems
\cite{takayanagi98,williams99} suggest a more practical alternative. 

Fig.\ref{fig:setup}(a) shows the simplest instance (Set-up I) of singlet Cooper pairs
tunneling from a superconducting lead through a point contact (PC) into the
 singlet FQH state which in turn is in ohmic contact with a  normal lead. Here we assume that the geometry of the FQH fluid is such that it is not significantly rearranged by the proximity of the superconductor.
Unlike tunneling of Cooper
pairs into fully polarized Laughlin states in Ref.\cite{fisher94},
the tunneling conductance into spin singlet FQH states would not be exponentially suppressed
since in this case pair-breaking
spin flip processes are not required.
In set-up II, we consider a QH bar geometry which
makes ohmic contact with two normal leads, and a tunnel contact with
a SC wedge which acts as
a third lead. 

Figures 1(b) and (c) show the Set-up II in two
different regimes of energy scale which depend on the experimental
parameters (for example, temperature and
bias voltage). In the high-voltage (``ultra-violet" (UV)) limit (Fig.1(b)), the allowed pair
tunneling process is effectively the same as in Set-up I. However, in the
low-voltage (``infra-red" (IR)) regime (Fig.1(c)) the presence of the SC lead induces backscattering processes among the edges of the QH bar which leads to a splitting of the FQH fluid into two droplets by quasiparticle tunneling\cite{kane92a}. 
We find that in this regime the most favored
pair tunneling process is the one in which pairs from the SC lead
split into the two droplets as entangled electrons. Hence, the low energy behavior of the junction is governed by a new type of fixed point, a
{\em quantum entangled fixed point} (QEFP), characterized by a boundary condition which mixes the phases of the spin channels of the split singlet FQH fluids and the SC into a singlet state.
 
In what follows, we present the  formalism for describing the FQH edge
states, the SC degrees of freedom and tunneling between the two
systems. In the SC lead the phase of the superconductor
can be pinned or endowed with its own dynamics, and we account
for both possibilities. We calculate the tunneling conductances
for the processes shown in Fig.\ref{fig:setup} and discuss their
behavior.

The starting point of all our analyses is a model Lagrangian for the
FQH-SC junction that describes the dynamics of the FQH edge and the SC
lead, and the tunneling between them through a single PC, given by 
\begin{equation}
\mathcal L = \mathcal L_0^{FQH}+\mathcal L_0^{SC}+\mathcal L_t.
\label{eq:tot-L}
\end{equation} 
We employ
the standard formalism of Ref.~\cite{wen95,milovanovic97,lopez01} for the singlet FQH edge
states and focus on the primary spin-singlet Halperin states \cite{halperin83} $(m,m,n)$ (with $n=m-1$, $m$ odd and filling factor $\nu_\pm=2/(2n\pm 1)$)
not only  for  the sake of simplicity but also since these states have been
observed experimentally \cite{leadley97,kronmuller98,cho98}, although these ideas can be extended to the Jain-like hierarchy of singlet FQH states \cite{lopez95,lopez01}. 
These edge states can
be described solely in terms of one propagating chiral boson per each
spin component or, equivalently, in terms of a charge chiral boson $\phi_c$ and a spin chiral
boson $\phi_s$. For the Halperin primary states with filling factor $\nu_\pm =2/(2n\pm 1)$, the resulting edge effective Lagrangian is
\begin{equation}
\begin{split}
{\mathcal  L}_0^{FQH}=&\frac{1}{4\pi}\partial_1\phi^c(\partial_0\phi^c-v_c\partial_1\phi^c)\\
+&\frac{1}{4\pi} \; \partial_1\phi^s( \pm \partial_0\phi^s-v_s\partial_1\phi^s).
\end{split}
\label{eq:edge-Lagrangian}
\end{equation}
The electron operator, constructed from quasiparticles to have correct quantum
numbers and statistics, takes the form
\begin{equation}
\psi^\dagger(x)_{e,\uparrow\downarrow}=\eta_{\uparrow,\downarrow}:\exp\left(\frac{i}{\sqrt{\nu}}\phi^c(x)\right):\times
:\exp\left(\pm\frac{i}{\sqrt{2}}\phi^s(x)\right):
\label{eq:electron-op}
\end{equation}
up to irrelevant operators~\cite{lopez01}, where
$\eta_{\uparrow,\downarrow}$ are Klein factors that ensure fermionic statistics.
From this we can use operator product expansions to obtain the creation operator for a charge $2e$
spin-singlet Cooper pair $\psi^\dagger_{2e}(x)$ having tunneled into the
FQH side: 
 \begin{equation}
\psi^\dagger_{2e}(x) = \frac{1}{\sqrt{2\pi}} \;
:\exp\left(i\frac{2}{\sqrt{\nu}}\phi^c(x)\right):
\label{eq:FQH-Cooper-pair}
\end{equation}
We now turn to the SC side and consider the
aforementioned possibilities of the superconducting phase being either
(i) pinned or (ii) dynamic. Whether or not the quantum dynamics of the superconductor needs to be
taken into account depends on the size of the lead, its coupling to the
environment, and the nature of the contact. 
In principle, the phase degree of freedom in the SC ought to exhibit
quantum mechanical fluctuations and should be treated dynamically. However,
equilibration with the external leads causes dissipation leading to the pinning of the phase of the superconductor \cite{caldeira-leggett83}.

When we take the phase fluctuation into consideration, we describe the
Cooper pair wave function in the  
superconducting side in terms of phase excitations with
charge $2e$. 
Reasoning along the lines of Ref. \cite{chamon97},  
one finds that only one channel for Cooper pairs couples to the PC
and that the Cooper pair operators can be described in terms of a single chiral 
boson $\Phi(x)$ on an infinite line
with Lagrangian density 
\begin{equation}
{\mathcal L}_0^{SC} = \frac{1}{4\pi}\partial_1\Phi(\partial_0\Phi-v_{sc}\partial_1\Phi)
\label{eq:SC-Lagrangian}
\end{equation} 
with the Cooper pair operator given in terms of $\Phi(x)$ by
\begin{equation}
\Psi_{2e}(x) = \frac{1}{\sqrt{2\pi}} \; :e^{i\sqrt{2}\Phi(x)}: 
\label{eq:SC-Cooper-pair}
\end{equation}
to have charge $2e$ and bosonic statistics.
On the other hand, when the coupling to the environment is significant
enough to pin the phase of the superconductor, the superconducting
order parameter acquires  a finite expectation value, and 
the Cooper pair operator $\Psi_{2e}$ on the superconducting side
can be replaced by its expectation value
$\langle\Psi_{2e}\rangle$. Then the SC effectively acts as
a reservoir of Cooper pairs that can tunnel into the FQH edge states
with the tunnel coupling proportional to $\langle\Psi_{2e}\rangle$.

Turning now to the tunneling contribution to the Lagrangian density,
all the processes shown in Fig.\ref{fig:setup} can be described by the tunneling density
\begin{equation}
{\mathcal L}_{t}=\Gamma_{ab}\delta(x)
e^{ie(V_{a}+V_{b})t/\hslash}(\psi_{a\uparrow}^{\dagger}
\psi_{b\downarrow}^{\dagger} -(\uparrow\leftrightarrow\downarrow))
\Psi_{2e} + h.c.,
\label{eq:SC-FQHtunneling}
\end{equation}
where $\psi^{\dagger}_{a j}$ is the electron creation operator
in edge state $a$ and with spin $j$. The incoming edge-state
$a$ is assumed to be held at a potential drop $V_{a}$
with respect to the superconductor and similarly for the edge state $b$. 
The tunneling events in Fig.\ref{fig:setup} are of two types-
tunneling $\Gamma$ wherein whole pairs tunnel into one
edge state, and $\tilde \Gamma$ where Cooper pairs split
into two edge states, {\it i. e.\/}
\begin{equation}
\Gamma\equiv\Gamma_{aa}; \quad
\tilde \Gamma\equiv\Gamma_{ab}, \quad a\neq b,
\label{eq:gammas}
\end{equation}
where for convenience, we have assumed that all tunneling
strengths for each kind of process are equal, but our results
can be generalized straightforwardly when this is not the case.
When $a=b$, the Cooper pair tunneling amounts to the creation of a
whole pair in one edge represented by the operator $\psi_{2e}^\dagger$ of
Eq.\eqref{eq:FQH-Cooper-pair} with  $e(V_{a}+V_{b})=2eV$, 
reflecting the fact that tunneling involves charge $2e$
Cooper pair processes. On the other hand, when $a\neq b$,
Eq.\eqref{eq:SC-FQHtunneling} describes the splitting of Cooper pairs
into two electrons tunneling into separate edge states, but retaining the spin-singlet phase correlations of the Cooper pair. 

In the case of Set-up I (Fig.\ref{fig:setup}a), the Cooper pair tunnelling amplitude (into their only available edge
state) is given by $\Gamma$. However, even for Set-up
II (Fig.\ref{fig:setup}b,c), the only significant tunneling process in the UV limit is from the SC to the closest
(say, right-moving) edge-state, once again described by the amplitude $\Gamma$. 
However, the presence of the SC lead
necessarily depletes the right-moving edge towards the left-moving
edge, providing a matrix element for inter-edge quasiparticle tunneling processes at the PC. 
Since such tunneling processes in a Hall bar are relevant perturbations, there exists a crossover energy scale 
 $\Delta_{\rm split}$ below which
quasiparticle
tunneling becomes dominant \cite{fendley9596}
resulting in the splitting of the QH liquid into two separate fluids (or droplets) `$1$' and `$2$'
at asymptotically low temperatures and voltages, shown in Fig.\ref{fig:setup}c. The particular advantage of this setup lies in the fact that
what plays the role of a back gate is a SC, into and out of
which Cooper pairs can tunnel. Now, very much as in the nanotube
set-up suggested in Ref. \cite{entanglement}, tunneling currents
into systems `$1$' and `$2$' each have two contributions: a)
 from Cooper pairs tunneling from the SC into either one
of the edges of the disconnected droplets (with amplitude $\Gamma$), and b) from pairs emerging
from the SC, splitting into entangled electrons which 
are each carried away by an edge state (with amplitude $\tilde \Gamma$).

Equipped with the bosonized Lagrangian given above, we can now
perturbatively calculate the
tunneling conductance for the various situations. 
In the case where the SC phase is pinned, the contribution $\mathcal
L_0^{SC}$ drops out of the Lagrangian density of Eq.\eqref{eq:tot-L}
and the Cooper pair operator on the SC side gets replaced by its
expectation value in $\mathcal L_t$.
When the SC dynamics is important, 
it is convenient to perform a rotation of bosonic
fields \cite{chamon97}. For instance, for
the Cooper pair tunneling into a single edge-state, the 
new Lagrangian takes the form \cite{chamon97}
\begin{equation}
\begin{split}
{\mathcal L}& =
\frac{1}{4\pi}\partial_1\varphi(\partial_0\varphi-\partial_1\varphi)+\frac{1}{4\pi}\partial_1\tilde\varphi(\partial_0\tilde\varphi-\partial_1\tilde\varphi)\\
&+\Gamma\delta(x)\cos\left(\sqrt{1+\frac{2}{\nu}}\; \left(\varphi(0,t)-\tilde\varphi(0,t)\right)\right),
\label{eq:dynamic-L-tot}
\end{split}
\end{equation}
where the fields $\varphi$ and $\tilde\varphi$ are related to the FQH edge
state charge degrees of freedom $\phi_c$ of Eq.\eqref{eq:edge-Lagrangian} and 
the SC degrees of freedom $\Phi$ of Eq.\eqref{eq:SC-Lagrangian} by an orthogonal transformation with angle $\tan \theta=(\sqrt{2}-\sqrt{\nu})/(\sqrt{2}+\sqrt{\nu})$.

The tunneling conductances can be calculated to lowest order in tunnel
amplitude \cite{kane92a,chamon97}. They show scaling behavior in $T$ and $V$ with an exponent $\gamma$,
\begin{equation}
G_t(V\!=\!0,T)\propto \left(\frac{T}{\Delta}\right)^{2\gamma}, \quad
G_t(V,T\!=\!0)\propto\left(\frac{V}{\Delta}\right)^{2\gamma} 
\label{eq:condlim}
\end{equation}
where 
$\Delta$ is a 
crossover energy scale, which depends on the process, below which the perturbative result Eq.\eqref{eq:condlim} applies. For
 Cooper-pair tunneling we set $\Delta=T_K$ and $\gamma=\alpha$, while  for split
Cooper-pair tunneling $\Delta=\tilde{T_K}$, $\gamma=\beta$, where
\begin{equation}
T_K \sim 
\Gamma
^{-1/\alpha},
\quad
{\tilde T_K} \sim
{\tilde \Gamma}
^{-1/\beta},
\label{eq:cross-over}
\end{equation}
The tunneling amplitudes $\Gamma$ and $\tilde \Gamma$ are defined in
Eq.\eqref{eq:gammas}. Here we assume a high-energy cutoff $\Lambda$ set by the Landau level
spacing. (For the case in which the SC phase mode is pinned, the tunneling amplitudes get renormalized
by a factor of $\langle\Psi_{2e}\rangle$.) The values of the exponents $\alpha$ and
$\beta$ depend on whether or not the phase field $\Phi$ of the superconductor is dynamical or pinned by dissipation.
We find $\alpha=1/\nu-1$, $\beta=\frac{1}{2}(1/\nu-1)$ for the
pinned case and $\alpha=2/\nu$, $\beta=1/\nu+1/2$ for the dynamical case.

We now discuss the tunneling conductances associated with the specific
cases shown in Fig. \ref{fig:setup} by applying the general form of
Eq.\eqref{eq:condlim}. The conductance arising from Cooper pair
tunneling between the SC lead and a single FQH edge 
(Fig.\ref{fig:setup}a and Fig.\ref{fig:setup}b) has the form
\begin{equation}
\!\!\!G_t(V,0) \propto 
\frac{2e^2}{h}
\left(\frac{V}{T_K}\right)^{2\alpha}
\!\!\!, \quad
G_t(0,T) \propto
\frac{2e^2}{h}
\left(\frac{T}{T_K}\right)^{2\alpha}
\end{equation}
Thus, for all values of $\nu$ and at low energy scales, a SQH-SC junction, with a
superconductor whose phase is pinned, has a larger conductance
compared to one with dynamics.
This is expected since the states on both sides of the junction will have a smaller overlap if the phase of the SC fluctuates than if it does not.
So, for instance, for the $\nu=2/3$ SQH state,
$2\alpha=1$ for the pinned case
and $2\alpha=6$ with dynamics, whereas for the $\nu=2/5$ SQH state, the exponents are $2\alpha=3$ and $2 \alpha=10$ respectively.
In contrast, the tunneling
conductance into polarized FQH states is  exponentially
suppressed as one of the constituent electrons of the 
Cooper pair has to flip its spin in order to tunnel into the FQH state.
Thus, the observation of a power law
conductance is a direct signature of a singlet FQH state.

The situation in Fig.\ref{fig:setup}c, wherein quasiparticle
tunneling leads to splitting up of the Hall droplet into two
disconnected fluids, presents the most intriguing
possibility. The tunneling conductances $G_t^1$ and $G_t^2$ associated with Cooper pairs
tunneling into edge-states `$1$' and `$2$' is
\begin{equation}
\begin{split}
G_t^a(V\!=\!0,T) &\sim
A_T\!\left(\frac{T}{T_{K}}\!\right)^{2\alpha}\!+\!B_T
\!\left(\frac{T}{T'_{K}}\!\right)^{2\beta}\\
G_t^a(V,T\!=\!0) &\sim
A_V\!\left(\frac{2V_{a}}{T_{K}}\!\right)^{2\alpha}
\!+\!B_V\!\left(\frac{V_{1}\!+\!V_{2}}{T'_{K}}\!\right)^{2\beta}
\end{split}
\label{eq:two}
\end{equation}
where $A_T$, $A_V$, $B_T$ and $B_V$ are constants.

~In Eq.\eqref{eq:two} we see that the conductance has two contributions: a) one due to pair tunneling as a whole into one edge-state,
with strength $\Gamma=\Gamma_{aa}, a=1,2$, and
associated with chemical potential $2eV_{a}$,
and b) another due to processes,  
with strength $\tilde \Gamma=\Gamma_{12}$ associated with chemical
potential $e(V_1+V_2)$ in Eq.\eqref{eq:SC-FQHtunneling}, in which the Cooper pair {\em splits} into its constituent electrons which tunnel separately into the two edge-states as entangled electrons. 
The first
contribution has the same form as the conductances described in 
Set-up I (Fig.\ref{fig:setup}a) and Set-up II (Fig.\ref{fig:setup}b), while the second contribution
corresponds to Set up II, Fig.\ref{fig:setup}c.
By comparing the exponents associated with the two
contributions in Eq. \eqref{eq:two}, we find that
regardless of whether the SC phase is pinned ($\alpha=\frac{1}{\nu}\!-\!1$
and $\beta=\frac{\alpha}{2}$) or dynamic ($\alpha=\frac{2}{\nu}$ and
$\beta= \frac{\alpha}{2}+\frac{1}{2}$) in all cases $\alpha > \beta$,  and hence at low energies
the entangled process {\em always} dominates. For $1 <\nu <2$ the allowed SQH states are hole-like and their edge states are prone to an edge reconstruction with a $\nu=2$ layer being formed. In this range of filling factors the QEFP 
is not accessible.  
  
What are the observable signatures of the QEFP? A direct indication for charge splitting would be the observation of an
indirectly driven tunneling current, {\it e.g.\/} a non-zero current $I^1_t$
when $V_1=0$ but $V_2\neq 0$. A more direct signature of singlet
entanglement requires the measurement of spin current
correlations across the two FQH edges.
At the QEFP there are delicate correlations in the spin  currents of the two split singlet FQH fluids which exhibit correlations similar to those in discussed in Ref. \cite{entanglement} (for a different physical system). 
If $\vec n_1$ and $\vec n_2$, with relative angle $\theta$, denote two axes of spin polarization, the correlation functions of the spin currents $I_{i,\pm n_i}^s$, with polarization state $\mid \pm \vec n_i\rangle$ ($i=1,2$) at the QEFP obey \cite{entanglement}
\begin{eqnarray}
\langle I^s_{1,\pm \vec n_1} I^s_{2,\pm \vec n_2}\rangle&=& e \sin^2(\theta/2)\; \langle I^s_{2,\vec n_2}\rangle
\nonumber \\
\langle I^s_{1,\pm \vec n_1} I^s_{2,\mp \vec n_2}\rangle&=& e \cos^2(\theta/2)\; \langle I^s_{2,\vec n_2}\rangle
\end{eqnarray}

In summary, we discussed the physics of a tunnel junction of
a singlet SC lead and a FQH state, and analyzed its transport properties.
We discussed the power-law behaviors of the tunneling conductances in temperature and voltage which should
 be observable {\em only} for singlet FQH states, thus filtering out
specific filling fractions and probing their spin polarization.
We found that the low energy regime of this system is governed by a QEFP in which the Hall bar is effectively split into two 
QH droplets into which Cooper pairs can split and tunnel coherently as {\em entangled electrons}.
The SQH/SC junction discussed here 
has 
a natural mechanism leading to a quantum entangled fixed point.    
The geometry of the pair splitting state presents analogies with
those of Ref. \cite{y-junction} for observing quasi-
particle and electron tunneling.
Finally, the set-up and measurements
proposed here can be used to investigate spin transitions in QH states,
as well as open issues on the Luttinger liquid nature of edge states.

We thank J. Eckestein and S. Flexner for 
discussions.
This work was supported in part by
the National Science Foundation grants No. DMR 01-32990 (EAK and EF), EIA01-21568, PHY00-98353 and by the Department of Energy contract No. DEFG02-96ER45434 (SV).


\end{document}